\documentclass[a4paper,12pt]{JHEP3}
\usepackage{amssymb}
\usepackage{mathrsfs}
\usepackage{cite}
\usepackage{texdraw}
\usepackage[english]{babel}
\usepackage{epsfig}

\csname @addtoreset\endcsname{equation}{section}

\newcommand{\bea}{\begin{eqnarray}}
\newcommand{\eea}{\end{eqnarray}}

\def\be{\beta}
\def\ga{\gamma}
\def\de{\delta}
\def\ep{\epsilon}

\def\Om{\Omega}

\def\fft#1#2{{#1 \over #2}}
\def\nn{\nonumber}

\title{Thermodynamics at the BPS bound for Black Holes in AdS}
\author{Pedro J. Silva \\
Dipartimento di Fisica dell'Universit\`a di Milano, and\\
INFN, Sezione di Milano\\
Via Celoria 16, I-20133 Milano, Italy \\
Institut de Fisica d'Altes Energies (IFAE),\\
E-08193 Bellaterra (Barcelona), Spain.\\
E-mail: \email{pedro.silva@mi.infn.it}}

\abstract{In this work we define a new limiting procedure that
extends the usual thermodynamics treatment of Black Hole physics,
to the supersymmetric regime. This procedure is inspired on
equivalent statistical mechanics derivations in the dual CFT
theory, where the BPS partition function at zero temperature is
obtained by a double scaling limit of temperature and the relevant
chemical potentials. In supergravity, the resulting partition
function depends on emergent generalized chemical potentials
conjugated to the different conserved charges of the BPS solitons.
With this new approach, studies on stability and phase transitions
of supersymmetric solutions are presented. We find stable and
unstable regimes with first order phase transitions, as suggested
by previous studies on free supersymmetric Yang Mills theory}


\begin{document}

\section{Introduction}\vspace{.5cm}

Classical Gravity shows many properties that are in fact
thermodynamical in its very nature like for example Black Hole
physics. In fact, General relativity (GR) can be thought as a
thermodynamic theory of space-time, emerging at low energy regimes
from a fundamental theory of space-time like string theory. This
framework suggests that GR could be understood as a mean-field
theory of thermodynamic nature.

Following this line of thoughts, it seems natural to approach
quantum gravity, studding the statistical behaviour of the
relevant ultraviolet degrees of freedom that in the infrared
condense to GR (i.e. the search for a rational foundation of GR).

The search for such ultra-violet partons is a complicated task,
and there is no clear answer of how to proceed. On the other hand,
there is little doubt that string theory contains a quantum theory
gravity, and delivers GR as an effective theory. The recurring
unsolved problem is our incapacity to solve string theory, and
hence understand the quantum gravity sector.

Within string theory, there are particularly simple systems where
dramatic simplifications occur due to the existence of symmetries
that effectively freezes almost all the degrees of freedom and
dynamics. In these cases, some understanding has been archived
resulting in the famous counting of microstates for a class of
supersymmetric Black Holes, or even some extreme but not
supersymmetric Black Holes
\cite{Strominger:1996sh,Khuri:1995xq,Emparan:2006it}.

The AdS/CFT duality \cite{M1,Gubser:1998bc} also has contribute to
improve our otherwise poor understanding of quantum gravity in the
case where there is a negative cosmological constant. Here, there
is a fully well defined quantum field theory ($SU(N)$ ${\cal N}=4$
SYM) that is equivalent to string theory with fixed asymptotic
conditions in a negative cosmological constant background. It
turns out that in the large $N$ limit, at strong 't Hooft
coupling, this field theory describes the GR sector.

In the literature there are many studies of quantum gravity and
its dual significance in the CFT theory. Again, up to now ${\cal
N}=4$ SYM theory at strong 't Hooft coupling is too complicated to
be fully comprehend and the program is far from completion.
Nevertheless, the statistical approach to GR has concrete
realization in studies on Black Holes in AdS, dual to finite
temperature versions of the CFT, or in studies in particularly
simple sectors of the CFT where due to symmetries progress is
archived. For example see
\cite{Silva:2005fa,Berenstein:2005aa,Balasubramanian:2005mg}.

Recently, there were found supersymmetric Black Hole solutions in
$AdS_5$ preserving only two real supercharges
\cite{Kunduri:2006ek,Gutowski:2004ez,Gutowski:2004yv,Chong:2005da,
Chong:2005hr,Chong:2006zx}. These Black Holes (BH) provide a new
arena were to search for microscopic signals of quantum gravity in
the AdS/CFT framework. It would be very interesting to derive the
entropy of such BH and in general understand its thermodynamical
properties from the CFT point of view.

In \cite{Kinney:2005ej}, it was defined an index to count states
in the CFT and also studied the statistical properties of the
different supersymmetric sectors in the free theory limit.
Unfortunately, the supersymmetric BH sector seems to be invisible
to this new index, while the statistical analysis suggested that
within the different BPS sectors there are complicated structures
with instabilities and phase transitions some of which are related
to Bose-Einstein condensations.

The aim of this work is to develop a framework where to study the
statistical properties of supersymmetric solitons in AdS. To do
this, we review some basic features of statistical mechanics with
particular focus on the $T=0$ limit where supersymmetry appears.
Then, as an example of the power of this new approach, we apply
these techniques to study supersymmetric BH solutions in AdS. In
the supersymmetric regime, we find a rich structure with phase
transitions in a well defined ensemble picture that matches the
canonical statistics analysis of dual CFT.

In section \ref{sec2}, we define the theoretical framework and the
particular procedure to obtain the thermodynamics at the BPS
bound. In section \ref{sec3}, we apply this procedure to the
rotating BTZ BH \cite{Banados:1992wn} and the supersymmetric
$AdS_5$ BH solutions found in \cite{Chong:2005hr}, as main
examples of the power of our new framework. In section \ref{sec4},
we begin the study on the stability and phase transitions of the
above supersymmetric BH. A more profound study of the
thermodynamic of the above BH with the corresponding comparison
with the CFT partition function is currently under
study\cite{futuro}. At last, in section \ref{sec5} we close with
an overview, and a discussion on other BH solutions suggesting
possible future directions

\section{Statistical mechanics and thermodynamics in GR}\vspace{.5cm}
\label{sec2}

Usually, the starting point of studies in statistical mechanics is
the definition of the micro canonical ensemble. Here, all the
physical states are label by a set of extensive quantities such as
energy $E$, angular momentum $J$, electric charge $Q$, etc. Then
it is assume that all states have the same probability to be
measure. The partition function $\Xi_{(E,J,Q)}$ is defined as the
total number of states in the ensemble, while the entropy $S$ is
defined as
\bea S=\ln \Xi_{(E,J,Q)}\quad\hbox{such that}\quad dS=\be dE+ \ga
dJ +\de dQ \,,\nonumber \eea
where $(\beta,\gamma,\delta)$ are the conjugate variables of
$(E,J,Q)$ respectively. In particular, note that the entropy
doesn't have to be zero at zero temperature, but in general is a
function of the other charges.

To study the properties of the system, sometimes it is useful to
exchange a few of the extensive variables with its conjugates
partners by means of Legendre transformations. Such
transformations have the physical interpretation of changing the
type of ensemble (provided these maps are well defined, all
ensembles are equivalent). In general, we have a grand canonical
ensemble when some of the charges are replaced by their conjugated
variables (here denoted in general as "potentials"). For example,
the form of the partition function $\Xi_{(\be,\ga,Q)}$ is
\bea \Xi_{(\be,\ga,Q)}=\sum_\nu e^{(-\be E_\nu -\ga J_\nu)}\,,
\nonumber \eea
where the sum is over all states $\nu$ of the ensemble with fixed
charge $Q$. The rational foundation of thermodynamics relies on
the identification of all the empirical thermodynamical
definitions like the different functions and laws, in terms of
statistical mechanics concepts. In particular, working in the
generalized grand canonical ensemble where all the charges have
been replaced by the potentials, the following identification is
obtained\footnote{The right-hand side of this equation corresponds
to the generalized Gibbs free energy divided by the temperature
$T$.}
\bea -\ln\Xi_{(\be,\Om,\Phi)}=\be E -\be\Om J - \be\Phi Q -S\,,
\nonumber \eea
where $\ga =-\be\Om$, $\de=-\be\Phi$ and $\be$ is the inverse of
temperature $T$.

Consider now the case, where our theory is equipped with
supersymmetry and we are interested to study its statistical
properties. First of all, note that supersymmetric states form a
subspace in the original Hilbert space of our theory. Therefore,
we have to constraint the initial partition function to this
supersymmetric hypersurface in order to account for the relevant
statistical properties. One of the key point we want to stress and
utilize along this work is that:
\\

{\it The supersymmetric partition function can be defined as a
combination of limits for the different potentials, but not as the
sole naive limit $\beta \rightarrow \infty$}.
\\

To make this clear, just note that all supersymmetric states are
annihilated by a given set of supercharges. Therefore these states
saturate a BPS inequality that translates into a series of
constraints between the different physical charges. For
definiteness, let us consider a simple example, where the BPS
bound corresponds to the constraint $E=J$. This type of BPS bound
appears in two dimensional supersymmetric models, like for example
the effective theory of $1/2$ BPS chiral primaries of $N=4$ SYM in
$R\otimes S^3$(see
\cite{Corley:2001zk,Berenstein:2004kk,Caldarelli:2004ig}). Then,
defining the left and right variables
\bea E^\pm= \hbox{$1\over2$}(E_\nu\pm J_\nu)\,,\quad\be_\pm=(\be
\pm \ga)\,, \nonumber \eea
$\Xi_{(\be,\ga,Q)}$ can be rewritten as 
\bea \Xi_{(\be,\ga,Q)}=\sum_\nu e^{-(\be_+E_+ + \be_-E_-)}\,.
\nonumber \eea
At this point, it is clear that taking the limit $\be_-\rightarrow
\infty$ while $\be_+\rightarrow \xi$ constant, gives the correct
supersymmetric partition function. The above limiting procedure
takes $T$ to zero, but also scales $\Om$ in such a way that the
new supersymmetric conjugated variable $\xi$ is finite and
arbitrary. Note that among all available states, only those that
satisfy the BPS bound are not suppress in the sum, resulting in
the supersymmetric partition function
\bea \Xi_{(\xi,Q)}=\sum_{\nu} e^{-\xi J}\,, \nonumber \eea
where the sum is over all supersymmetric states $\nu$ with $E=J$
at fixed charge $Q$ \footnote{We thank J. Maldacena for clarifying
this procedure in the dual CFT in relation with
\cite{Kinney:2005ej}.}. The above argument shows how to find the
relations between statistical mechanics and thermodynamics for
supersymmetric configuration as a multi-scaling limit.

On the other hand, BH physics inspired a body of definitions and
equations that effectively defines a thermodynamic theory in terms
of space-time variables \cite{Bekenstein:1973ur}. The missing step
in this "thermodynamics of space-time" is its rational foundation
in terms of quantum gravity\footnote{see \cite{Silva:2005fa} for
some results within AdS/CFT}. In more detail, it has been shown
that a semi classical approximation of the "quantum gravity
partition function" results in the exponential of the Euclidean
supergravity action $I$, evaluated on the corresponding solitonic
solution\footnote{The actual evaluation of $I$ and other
space-time quantities like energy, entropy, electric charges, etc
is a delicate task, where careful regularization have to be done.
Here we assume that there is always a way to carry on such
procedures.}. The Euclidean action holds unexpected properties
that interrelates it with other space-time variables such as
Hawking temperature, energy, entropy, angular momentum, etc that
defines the thermodynamics of space-time. In fact, $I$ plays the
role of free energy divided by temperature and hence satisfies the
equation
\bea I=\be E-\be\Om J -\be\Phi Q-S\,, \label{QSR}\eea
sometimes called the "quantum statistical relation" (QSR), where
all quantities are evaluated on the particular solitonic solution
of interest.

We point out that this thermodynamics is not only applicable to BH
solutions, but seems to carry on to other solitonic solutions.
Therefore we conjecture and use as a working hypothesis that:
\\

{\it the BH thermodynamics of space-time extends naturally to a
general principle of space-time physics}.
\\

In fact, in \cite{Jacobson:1995ab} the very local GR equations
where derived as equation of state of a generalization of BH
thermodynamics that involves causal horizons. It is known, that
the first law of BH thermodynamics is satisfied by a series of
topological solitons that do not look like a BH
\cite{Cvetic:2005zi}. To be more precise, we are thinking to
include all solitons with any temperature, even zero (like extreme
and supersymmetric solutions BH), topological solitons (where
there is no horizon at all!), etc.

In this work we will apply the above conjecture to supersymmetric
BH solutions in AdS and its non-supersymmetric extensions
\cite{Chong:2005hr}. The reason behind this pragmatic approach is
to open up a new mechanism to investigate on $1/2,1/4,1/8$ and
$1/16$ supersymmetric configurations in the AdS/CFT duality and
try to find traces of a microscopic description of the QSR.

Since we want to study supersymmetric solitons, the first task we
face is how to recover the $T=0$ thermodynamics in the GR
framework. Note that in the supersymmetric limit, the
thermodynamic potentials are independent of physical observables
(like charge, energy, etc). For example, in a BPS rotating BH, the
angular velocity of the horizon $\Om$, equals the velocity of
light regardless its mass or angular
momentum\cite{Hawking:1998kw}. The same happens to the
electrostatic potential $\Phi$ that also can be understood as
angular velocity once the solution is uplift to ten or eleven
dimensions. Therefore, for BPS solutions, we do have some
thermodynamical variables at our disposal, like entropy and
conserved charges, but we seem to lack of thermodynamical
potentials.

Nevertheless, in our previous discussions of supersymmetric limits
in statistical mechanics, we learned that the correct procedure is
to consider a combined limit of the different off-BPS potentials.
Accordingly, taking the QSR equation it is easy to see that if the
charges behave as
\bea E \rightarrow E_{bps}+ O(\be^{-2})\,, \quad J\rightarrow
J_{bps}+ O(\be^{-2})\,,\quad Q\rightarrow Q_{bps}+ O(\be^{-2})\,
\label{limit1}\eea
things will not work correctly unless the GR potentials behave
as\footnote{Actually, what is need is that the different
potentials scale as a fine part plus a part the goes like the
inverse of the leading part of $\beta$. For BH, as we approach the
extreme limit, $\be$ tends to infinite, but this is not the
general case for solitons that include BPS regimes.}
\bea \be \rightarrow \infty\,, \quad \Om \rightarrow
\Om_{bps}-{w\over\be }+ O(\be^{-2})\,,\quad \Phi \rightarrow
\Phi_{bps}-{\phi\over\be }+ O(\be^{-2})\,, \label{limit2} \eea
where the "$bps$" subscript defines the corresponding
supersymmetric values. If the above general behaviour is not
archived by the soliton under study, the supersymmetric limit will
no be recover, showing a breakdown of our conjecture on the
general nature of the thermodynamical properties of space-time or
a failure of the GR description of the corresponding
soliton\footnote{We recalled that GR is an effective theory and we
may very well need to take into account string corrections to the
metric. For example, it is believed that the superstar solution
\cite{Behrndt:1998ns} receives stringy corrections that should
dress its naked singularity with an appropriated event horizon.}.

Assuming that the above relations hold, we get the following
expression
\bea I=
\be\left(E_{bps}-\Om_{bps}J_{bps}-\Phi_{bps}Q_{bps}\right)+w
J_{bps}+\phi Q_{bps}-S_{bps}+O(\be^{-1})\,. \nonumber \eea
since the first term multiplying $\beta$ is identically zero due
to the BPS property of the limiting solution, we end up with the
"supersymmetric version of the the quantum statistical relation"
(SQSR)
\bea I_{bps}=w J_{bps}+\phi Q_{bps}-S_{bps}\, ,\nonumber \eea

The above definitions of supersymmetric potentials and the SQSR
are a concrete realization of our new perspective in the BH
thermodynamics. Armed with these new objects, we can study the
different thermodynamical properties of BPS solitons and compare
them with the dual statistical behaviour.

In the following section, we will check that the above framework
is verified and that indeed the BPS potentials and the SQSR are
well defined for AdS solitons. Then, we start the study of its
thermodynamical consequences, like stability and phase
transitions.

\section{Thermodynamics of supersymmetric solitons in AdS}
\label{sec3}

The aim of this section is to test the general properties derived
before. In the long run, we are interested in supersymmetric
solutions of five dimensional gauge supergravity because these
solutions can be also studied via its holographic relation to N=4
SYM theory extending studies like \cite{Kinney:2005ej}.
Supersymmetric BH solutions in $AdS_5$ were found first in
\cite{Gutowski:2004ez,Gutowski:2004yv} and then in
\cite{Chong:2005da,Chong:2005hr,Chong:2006zx}. Recently, in
\cite{Kunduri:2006ek} all the families were generalized to a
single solution. Unfortunately, only the solutions of
\cite{Chong:2005da} and those of \cite{Chong:2006zx} are known in
an off-BPS regime. There are other $AdS_5$ BH solution like
\cite{Behrndt:1998ns,Behrndt:1998jd,Klemm:2000gh} that have ill
defined BPS limits. In these later cases, the BPS limit degenerate
to naked singular solutions that are believe to get stringy
corrections, that in turn should cure this behaviour either
dressing the singularity with an event horizon or smoothing it
out. Also, in a different framework, we have the BTZ BH solution
\cite{Banados:1992wn}, that has a well defined supersymmetric
limit and has been extensively studied in the $AdS_3/CFT_2$
correspondence.

Our strategy is plain and simple. Given a solution that contains a
well defined BPS regime, we find an adequate description of its
off-BPS character. Then, we expand around it to recover the SQSR
and the corresponding supersymmetric potentials. At this point, we
should be ready to investigate its statistical dynamical
properties like phase transitions and instabilities.

In this work, we will deal with BH solutions that have well
behaved supersymmetric limits, like the BTZ BH and the solutions
of \cite{Chong:2005hr}.

\subsection{Statistical properties of rotating BTZ Black Hole at T=0}

As a first example, we start our studies with the rotating BTZ BH
solution \cite{Banados:1992wn}. This BH appears naturally as the
near horizon geometry of a system of $Q_1$ D1-branes, $Q_5$
D5-branes and a wave with momentum $N$. Therefore lies within the
$AdS_3/CFT_2$ correspondence. The rotating BTZ metric can be
written as
\bea ds^2 =
-\fft{(r^2-r_+^2)(r^2-r_-^2)}{r^2}\,dt^2+\fft{r^2l^2}{(r^2-r_+^2)(r^2-r_-^2)}\,dr^2
+ r^2(d\phi+\fft{r_+r_-}{r^2}\,dt)\,. \nonumber\eea
The energy $E$ and angular momentum $J$ are given by
\bea E=\fft{k}{2l^2}(r_+^2+r_-^2)\quad\hbox{and}\quad
J=\fft{k}{l^2}(r_+r_-)\,, \nonumber\eea
where $k=Q_1Q_5$ is used since equals the level of the dual
Kac-Moody superconformal algebra and $l$, the radius of $AdS_3$ is
also a function of $Q_1\,,Q_5$, the string coupling constant and
the compactified four dimensional volume\footnote{We follow the
same notation and conventions of \cite{Maldacena:1998bw}, that we
recommend for further reading in the $AdS_3/CFT_2$ duality.}. The
above BH has an internal horizon at $r_-$ and external horizon at
$r_+$. The supersymmetric limit is recovered when the energy
equals the angular momentum, resulting in the collapse of both
horizons i.e. $r_+=r_-$.

In \cite{Maldacena:1998bw}, a very convenient description is given
in terms of the left and right temperatures $(T_+,T_-)$ of the
dual CFT theory, such that $r_\pm=\pi l(T_+\pm T_-)$. Then all the
relevant thermodynamic potentials like the Hawing temperature $T$,
angular velocity of the external horizon $\Om$, entropy $S$ and
charges $E$ and $J$, can be written in terms of $T_+$ and $T_-$ as
follows
\bea &\be\equiv{1\over T}=({1\over 2T_+}+{1\over 2T_-})\,,\quad
\Om=\left(\fft{T_+-T_-}{T_++T_-} \right)\,, \quad
E=\fft{\pi^2k}{2l^2}(T_+^2+T_-^2)\,,& \quad
\nonumber \\
&J=\fft{\pi^2k}{2l^2}(T_+^2-T_-^2)\,,\quad S=2\pi^2k(T_++T_-)\,.
\nonumber \eea

We are interested in the $T_-=0$ limit of the above soliton and in
particular in the corresponding QSR equation \ref{QSR}. The
expansion in terms of the off-BPS parameter $T_-$ of the relevant
thermodynamic variables gives
\bea &\be\rightarrow {1\over 2T_-}+{1\over 2T_+}\,,\quad
\Om\rightarrow 1-\fft{2}{T_+}T_-+O(T_-^2)\,,\quad S\rightarrow
2\pi^2kT_+ + O(T_-^1)\,,&\nonumber\\
&E\rightarrow \pi^2kT_+^2 +O(T_-^2)\,,\quad J\rightarrow
\pi^2kT_+^2 +O(T_-^2)\,.& \nonumber \eea

The above expansion defines the BPS value for the $E$, $J$ and $S$
as $E_{bps}=J_{bps}=\pi^2kT_+^2$ and $S_{bps}=2\pi^2kT_+$. Also as
explain in the previous section, the BPS value of $\Om$ is
independent of the supersymmetric BH parameter $T_+$ and evaluates
to 1. Collecting term together for the QSR we get that
\bea I= \left({1\over 2T_-}\right)\left(E_{bps}-J_{bps}\right) +
\left({1\over 2T_+}\right)\left(E_{bps}-J_{bps}\right) +
\left({1\over T_+}\right)J_{bps}-S_{bps}+O(T_-^1)\,,\nonumber\eea
from with we deduce that the finite part $I_{bps}$, corresponding
to the BPS action satisfies the SQSR equation
\bea I_{bps}&&=wJ_{bps}-S_{bps}\,,\nonumber \\
&&=-\pi^2kT_+\,, \nonumber \eea
where $w=1/T_+$. Note that the action $I_{bps}$ is a negative
function for all $T_+>0$ showing stability and no phase
transitions with a very simple behaviour.

Nevertheless, this first example shows that in fact the
theoretical framework considered in the previous section, has an
explicit realization in the rotating BTZ BH. In principle, there
is no a priory reason why the supergravity solution has to follow
an expansion like (\ref{limit1}) and (\ref{limit2}). This is a non
trivial test on our conjecture for the general nature of the
thermodynamics of space-time.

Also, in this particular case it is remarkable that the
thermodynamic potential $w$ matches exactly the inverse of the CFT
dual left temperature $T_+$, in a striking parallelism to the CFT
statistical description of this supersymmetric sector (again, see
\cite{Maldacena:1998bw} for the dual CFT description).

\subsection{Statistical properties of $AdS_5$ Black Holes at T=0}
\vspace{.3cm}

\noindent Let us consider next, solitons on $AdS_5$ that are dual
to $SU(N)$ $\mathcal{N}=4$ SYM in four dimensions. The solution we
consider here was first presented in \cite{Chong:2005da}. In the
BPS regimen, these solutions preserved only a fraction of $1/16$
of the total $32$ supercharges, and depending on the different
range of values of its parameter space, describe BPS BH or
topological solutions with no horizon.

In general, the solution comes with two independent angular
momenta $(J_1,J_2)$, and a single electric charge $Q$
\footnote{This is a solution of minimal five dimensional gauge
supergravity.}. In terms of Boyer-Lindquist type coordinates
$x^\mu= (t, r, \theta, \phi,\psi)$ that are asymptotically static
(i.e. the coordinate frame is non-rotating at infinity), the
metric and gauge potential are
\bea ds^2 &=& -\fft{\Delta_\theta\, [(1+g^2 r^2)\rho^2 dt + 2q
\nu] \, dt}{\Xi_a\, \Xi_b \, \rho^2} + \fft{2q\,
\nu\omega}{\rho^2}\nn + \fft{f}{\rho^4}\Big(\fft{\Delta_\theta \,
dt}{\Xi_a\Xi_b} - \omega\Big)^2 + \fft{\rho^2 dr^2}{\Delta_r} + \nn \\
&&\quad\,\quad+\fft{\rho^2 d\theta^2}{\Delta_\theta}
+\fft{r^2+a^2}{\Xi_a}\sin^2\theta d\phi^2 +
      \fft{r^2+b^2}{\Xi_b} \cos^2\theta d\psi^2\,, \nonumber \\
A &=& \fft{\sqrt3 q}{\rho^2}\,
         \Big(\fft{\Delta_\theta\, dt}{\Xi_a\, \Xi_b}
       - \omega\Big)\,, \nonumber \eea
where
\bea \Delta_r &=& \fft{(r^2+a^2)(r^2+b^2)(1+g^2 r^2) + q^2 +2ab
q}{r^2} - 2m
\,,\nn\\
\Delta_\theta &=& 1 - a^2 g^2 \cos^2\theta -b^2 g^2
\sin^2\theta\,,\quad f= 2 m \rho^2 - q^2 + 2 a b q g^2 \rho^2\,,\nn\\
\rho^2 &=& r^2 + a^2 \cos^2\theta + b^2 \sin^2\theta\,,\quad \Xi_a
=1-a^2 g^2\,,\quad \Xi_b = 1-b^2 g^2\,,\nn\\
\nu &=& b\sin^2\theta d\phi + a\cos^2\theta d\psi\,,\quad \omega =
a\sin^2\theta \fft{d\phi}{\Xi_a} + b\cos^2\theta
\fft{d\psi}{\Xi_b}\,.\nn \eea
The relevant thermodynamical potentials are
\bea \be = \fft{2\pi r_+\, [(r_+^2+a^2)(r_+^2+b^2) +
abq]}{r_+^4[(1+ g^2(2r_+^2 + a^2+b^2)] -(ab + q)^2}\,,\quad \Phi=
\fft{\sqrt{3}qr_+^2}{(r_+^2+a^2)(r_+^2+b^2+abq)}\,,\nn\\
\Omega_1 = \fft{a(r_+^2+ b^2)(1+g^2 r_+^2) + b q}{
               (r_+^2+a^2)(r_+^2+b^2)  + ab q}\,,\quad
\Omega_2 = \fft{b(r_+^2+ a^2)(1+g^2 r_+^2) + a q}{
               (r_+^2+a^2)(r_+^2+b^2)  + ab q}\,.\quad\quad \nonumber
\eea
while the conserved charges are
\bea E= \fft{m\pi(2\Xi_a+2\Xi_b - \Xi_a\, \Xi_b) + 2\pi q a b
g^2(\Xi_a+\Xi_b)}{4 \Xi_a^2\, \Xi_b^2}\,,\quad  Q = \fft{\sqrt3\,
\pi q}{4 \Xi_a\, \Xi_b}\,, \nn\\
J_a = \fft{\pi[2am + qb(1+a^2 g^2) ]}{4 \Xi_a^2\, \Xi_b}\,,\quad
J_b = \fft{\pi[2bm + qa(1+b^2 g^2) ]}{4 \Xi_b^2\,
\Xi_a}\,.\quad\nonumber \eea
Finally the entropy is given by
\bea S=\fft{\pi^2 [(r_+^2 +a^2)(r_+^2 + b^2) +a b q]}{2\Xi_a \Xi_b
r_+} \,. \nonumber \eea
In all the above expressions, $r_+$ is largest positive root of
$\Delta_r=0$. Also for convenience, in the rest of the paper we
will set the $AdS$ radius to 1 i.e. $g=1$.

\noindent The BPS limit is achieved if
\[E - J_a -J_b - \sqrt3\, Q =0\]
that in terms of the for parameters $(m,q,a,b)$ translates into
$q(1+a+b)=m$.

As we said before, in the BPS limit, we not only have BH
solutions, but topological solutions, naked singular solutions,
and even over-charge and over-rotating solutions containing
pathologies like closed time-like curves. To define the correct
BPS limit we have to avoid the forbidden regions in the moduli
space of the solutions. At the end, in the BPS regime there are
two types of regular solitons, BH and topological solutions.

\vspace{.3cm} \noindent {\bf BPS BH solutions}\vspace{.3cm}

\noindent We have found that the following procedure does take us
safely to the BPS BH solutions. Define the off-BPS parameter $\mu$
such that $m=m_{bps}+\mu\quad\hbox{where}\quad m_{bps}=q((1+a+b)$.
To avoid the over-charged regimes that produce pathological
solutions with CTC, is enough to impose the constraint
$q=(1+a)(1+b)(a+b)$. As a result of the above, we have reduced the
number of independent parameters from the original four
$(m,q,a,b)$ to three $(\mu,a,b)$ out of which $\mu$ controls the
off-BPS nature of the solution. Summarizing we have 
\[ m=q(1+a+b)+\mu\,,\quad q=(1+a)(1+b)(a+b)\,.\]
With the above parametrization is straight forward to expand all
the thermodynamic quantities in terms of $\mu$, obtaining for the
potentials 
\bea \be&=&\left(\fft{\pi q\sqrt{r_{bps}^2+(1+a+b)^2}}{\sqrt{2}
(3r_{bps}^2+1+a^2+b^2)}\right)\fft{1}{\sqrt{\mu}}+ O(0)\,,\nonumber \\
\Phi&=&\sqrt{3}+\left(\fft{\sqrt{6}(ab-1)}{(1+a)(1+b)\sqrt{[r_{bps}^2+
(1+a+b)^2]r_{bps}^2}}\right)\sqrt{\mu}+O(1)\,,\nonumber \\
\Om_1&=&1+\left(\fft{\sqrt{2}(a-1)(a+2ab+b^2+2b)}{q\sqrt{[r_{bps}^2+
(1+a+b)^2]r_{bps}^2}}\right)\sqrt{\mu}+O(1)\,,\nonumber \\
\Om_2&=&1+\left(\fft{\sqrt{2}(b-1)(b+2ab+a^2+2a)}{q\sqrt{[r_{bps}^2+
(1+a+b)^2]r_{bps}^2}}\right)\sqrt{\mu} +O(1)\,,\nonumber \eea
 and for the conserved charges and entropy
\bea E= E_{bps}+O(1)\,,\quad J^1=J_{bps}^1+O(1)\,,\quad
J^2=J_{bps}^2+O(1)\,,\nonumber \\
Q=Q_{bps}\,,\quad S=S_{bps}+O(1)\,,\nn \qquad\qquad\qquad
\nonumber \eea 
where
\bea E_{bps}&=&\fft{\pi (a + b)(1-a)(1-b)+(1
+a)(1+b)(2-a-b)}{4(1 -a)^2 (1-b)^2}\,, \nonumber \\
J_{bps}^1&=&\fft{\pi (a +b) (2a +b + a b)}{4 (1-a )^2
(1-b)}\,,\quad J_{bps}^2=\fft{\pi (a +b) (a + 2b +
ab)}{4(1-a)(1-b)^2}\,,\nonumber \\
&&Q_{bps}=\fft{\sqrt{3}\pi(a+b)}{4(1-a)(1-b)}\,,\quad
S_{bps}=\fft{\pi^2(a +b)r_0}{2(1-a)(1- b)}\,, \nonumber \eea
with $r_{bps}^2=a+b+ab$, corresponding to the position of the BPS
horizon.

Evaluating the above expansion into the QSR, we get
\bea I_{bps}=\phi
Q_{bps}+w_1J_{bps}^1+w_2J_{bps}^2-S_{bps}\nonumber \eea where
\bea
w_1=\fft{\pi(1-a)(a+2ab+b^2+2b)}{r_{bps}(3r_{bps}^2+1+a^2+b^2)}\,,\quad
w_2=\fft{\pi(1-b)(b+2ab+a^2+2a)}{r_{bps}(3r_{bps}^2+1+a^2+b^2)}\,,\nonumber \\
\phi=\fft{\pi\sqrt{3}(a+b)(1-ab)}{r_{bps}(3r_{bps}^2+1+a^2+b^2)}\,.
\qquad\qquad\qquad\qquad\nonumber \eea

Let us consider what we have archived up to now. In first place,
the expansion in the off-BPS parameter $\mu$ has reproduce the
exact behaviour, anticipated in the general discussion, that
defines the SQSR for supersymmetric solitons. This is not a
trivial fact, since there is no reason a priory why things should
work as they do. It is simply another surprise of GR and another
confirmation of its thermodynamical nature. Second, in doing the
expansion, we have obtained the definition of the corresponding
supersymmetric potentials $(\phi,w_1,w_2)$. Naturally these
potentials comes as functions of the parameters $(a,b)$, mimicking
exactly the usual statistical dynamics derivation in the CFT.
These potentials are the variables that in the thermodynamical
sense, define the Generalized Grand canonical ensemble at zero
temperature. Third, with this new framework, we are able to study
the stability and phase transitions of the above solutions.
\\

\vspace{.3cm} \noindent {\bf BPS topological
solutions}\vspace{.3cm}

\noindent For the topological soliton sector, we have found that
the following procedure does take us safely to the BPS regime. As
before, define the off-BPS parameter $\mu$ such that
$m=m_{bps}+\mu\quad\hbox{where}\quad m_{bps}=q((1+a+b)$ but now,
to avoid the over-charged regimes that produce pathological
solutions with CTC, is enough to impose the constraint
$q=-(1+a)(1+b)(a+2b+ab)(b+2a+ab)$. This choice of constraints,
should be accompanied with the coordinate transformation
$R=r^2-r_{bps}^2$, where $\Delta_r(r_{bps})=0$, since
$r_{bps}^2=-(a+b+ab)^2$. As a result of the above, we have reduced
the number of independent parameters from the original four
$(m,q,a,b)$ to three $(\mu,a,b)$ out of which $\mu$ controls the
off-BPS nature of the solution. Summarizing we have 
\[ m=q(1+a+b)+\mu\,,\quad q=-(1+a)(1+b)(a+2b+ab)(b+2a+ab)\,.\]

Next, we expand all the thermodynamic quantities in terms of
$\mu$, obtaining for the potentials 
\bea
\be&=&\left(\fft{2\pi(2r_{bps}^2+a^2+b^2)}{r_{bps}^3(1+a+b+2r_{bps}^2)}\right)\mu
+
 O(2)\,,\nonumber\\
\Phi&=&\left(
\fft{\sqrt{3}qr_{bps}^2}{(1+a+b+2r_{bps}^2)}\right)\fft{1}{\mu}+
\fft{\sqrt{3}q}{(1+a+b+2r_{bps}^2)}+O(1)\,,\nonumber\\
\Om_1&=&\left(\fft{a(r_{bps}^2+b^2)(1+r_{bps}^2)}{(1+a+b+2r_{bps}^2)}\right)
\fft{1}{\mu}+\fft{a(2r_{bps}^2+1+b^2)}{(1+a+b+2r_{bps}^2)}+O(1)\,, \nonumber\\
\Om_1&=&\left(\fft{b(r_{bps}^2+a^2)(1+r_{bps}^2)}{(1+a+b+2r_{bps}^2)}\right)
\fft{1}{\mu}+\fft{b(2r_{bps}^2+1+a^2)}{(1+a+b+2r_{bps}^2)}+O(1)\,,\nonumber\eea
for the conserved charges
\bea E= E_{bps}+O(1)\,,\quad J^1=J_{bps}^1+O(1)\,,\quad
J^2=J_{bps}^2+O(1)\,,\quad Q=Q_{bps}\,,\nn \eea 
where
\bea &&J_{bps}^1=-\fft{\pi(2b+a+ab)(2a+b+ab)^2}{4(1-a)^2(1-b)}
\,,\quad J_{bps}^2=-\fft{\pi(2b+a+ab)^2(2a+b+ab)}{4(1-a)(1-b)^2}\,,\nonumber\\
&&
Q_{bps}=-\fft{\sqrt{3}\pi(2b+a+ab)(2a+b+ab)}{(1-a)(1-b)}\,,\quad
E_{bps}=\sqrt{3}Q_{bps}+J_{bps}^1+J_{bps}^2 \nonumber\eea
and finally the entropy S gives
\bea
S=\left(\fft{\pi(2r_{bps}^2+a^2+b^2)}{2r_{bps}(1-a^2)(1-b^2)}\right)\mu
+ O(2)\,.\nonumber \eea
Note the strange responds of all the thermodynamic functions to
the off-BPS expansion. $\be$ goes to zero while all the potential
diverge, but in such a way the the physical quantity $\be\Om$ or
$\be\Phi$ has a finite value. Although this behaviour seems
counterintuitive, we point out that these specific combinations of
$\be\Om$ and the others, are the physical periods of the angular
variables in the Euclidian regime (see section three in
\cite{Hawking:1998kw}), and therefore is reasonable kept them
constant along the expansion.

Evaluating the above off-BPS expansion into the QSR, we get
\bea I_{bps}=\phi Q_{bps}+w_1J_{bps}^1+w_2J_{bps}^2\,,\nonumber
\eea
where
\bea w_1=-\fft{2\pi
a(r_{bps}^2+b^2)(r_{bps}^2+1)}{r_{bps}^3(1+a^2+b^2+2r_{bps}^2)}\,,\quad
w_2=-\fft{2\pi
b(r_{bps}^2+a^2)(r_{bps}^2+1)}{r_{bps}^3(1+a^2+b^2+2r_{bps}^2)}\,,\nonumber\\
\phi=-\fft{2\sqrt{3}\pi
q}{r_{bps}^3(1+a^2+b^2+2r_{bps}^2)}\,.\qquad\qquad\qquad\qquad
\nonumber\eea

Unfortunately this is not the end of the story, since in the above
solution; we still have to impose another constraint to avoid a
conical singularity,
\bea 3(1+b)(2+b)a^2+2(4b^2+7b+1)a+(5b+1)=0\,. \nonumber\eea
This equation can be easily solved for $b$, to give $a=-b/(2+b)$
or $ a=-(5b+1)/(3+3b$. The firs option gives pure AdS space, so we
concentrate in the second option. The evaluation of the different
charges and potential is not difficult but tedious, here we the
final results first for the charges,
\bea J_{bps}^1=\fft{(b^2+4b+1)^2}{48(2b+1)^2}\,,\quad
J_{bps}^2=\fft{(b^2+4b+1)}{72(2b+1)}\,,\quad
Q_{bps}=-\fft{\sqrt{3}(b^2+4b+1)}{24(2b+1)}\,,\nonumber \\
E_{bps}=\sqrt{3}Q_{bps}+J_{bps}^1+J_{bps}^2\,,
\qquad\qquad\qquad\qquad\qquad\nonumber \eea
and for the potentials,
\bea
w_1&=&\fft{8(5b+1)(1+b)(5b^2-4b-1)(b^2+b-2)}{(b^4-6b^3+17b^2+16b+8)(2b+1)^3}\,,
\nonumber \\
w_2&=&\fft{96b^2(b^2+b-2)(b^3+3b^2-3b-1)}{(b^4-6b^3+17b^2+16b+8)(2b+1)^3}\,,
\nonumber \\
\phi &=&\fft{8\sqrt{3}\pi(1-b)^2(1+4b+b^2)}{(1+b)(2b+1)^3}\,.
\nonumber \eea

Again, as in the previous case we have found a finite expression
for the SQRS in terms of the supersymmetric charges and the
conjugated potentials. Note that as expected, in this case there
is no entropy. This solutions are in this sense like the regular
LLM solutions of the $1/2$ BPS sector \cite{LLM}, where there is a
well defined ensemble of chiral primaries exited, that do not
produce a sizeable entropy for regular solutions (see
\cite{Suryanarayana:2004ig} for more comments).

\section{Stability and Phase transitions}
\label{sec4}

The definition of the SQSR permits the study of the semi classical
partition function, as we vary the different chemical potentials
depending on the case of study. Like in the dual conformal field
theory at T=0, we have a rich physical structure with phase
transitions where the BH soliton is not any more the dominant
soliton, as in fact occurs in the dual CFT for $1/16$
supersymmetric sectors\footnote{In \cite{Kinney:2005ej}, the
supersymmetric partition function is studied in the free case,
using the CFT picture, while in the strong coupling limit, is
studied at low energies using the approximation of a gas of
supergravitons in AdS and at higher energies using the BH
solutions. In that analysis, phase transitions were found
explicitly in the free CFT theory.}.

Before presenting the stability analysis, it is important to
realized that these solitons are not the most general $1/16$
supersymmetric solutions. In the dual CFT, general states in the
$1/16$ supersymmetric representation depend of three R-charges and
two angular momenta. Even if we look for states with the same
R-charges, there is no need of an extra constraint relating the
two angular momenta and the electric charge. On the other hand,
for five dimensional gauge supergravity with 32 real supercharges,
we should have three different electric charges plus two angular
momenta, giving a grand total of five independent degrees of
freedom. In the supergravity soliton, all the electric charges are
collapsed into one\footnote{We are working in minimal gauge
supergravity, solutions with general different three electric
charges and two angular momenta are known, but only in the BPS
limit, and present a constraint to avoid un-physical solutions.}
and on the top of this, we have to impose a relation between the
electric charge and two angular momenta, to avoid CTC or naked
singularities. Therefore, it is reasonable to expect that there
should be more general solutions waiting to be discovered.

Keeping the above fact in mind, we proceed to study this solution,
that are the best we can do to scan the physical structure in the
$1/16$ BPS sector from the supergravity point of view. Basically
we are working in a constraint hypersurface of the full space of
supersymmetric BH solutions.

\subsection{\bf BPS BH solutions}\vspace{.3cm}

Evaluation of the $I_{bps}$ in this case gives
\bea I_{bps}=\fft{\pi^2(a+b)^2[-1+2b+b^2+b^2+a(2+5b+b^2)+
a^2(1+b)]}{4(1-a)(1-b)\sqrt{a+b+ab}(1+a^2+b^2+3(a+b+ab))}\,,
\nonumber \eea
where the range of the parameters $(a,b)$ is obtained from the
physical condition that $r_{bps}^2\geq 0$ or equivalently
$a+b+ab\geq 0$,$(a,b)\leq1$. The first inequality is the condition
that the position of the horizon is well defined and the second
comes from our normalization of the AdS radius.

In figure (fig. 1), we show a three dimensional plot of $I_{bps}$
as a function of $(a,b)$.
\EPSFIGURE{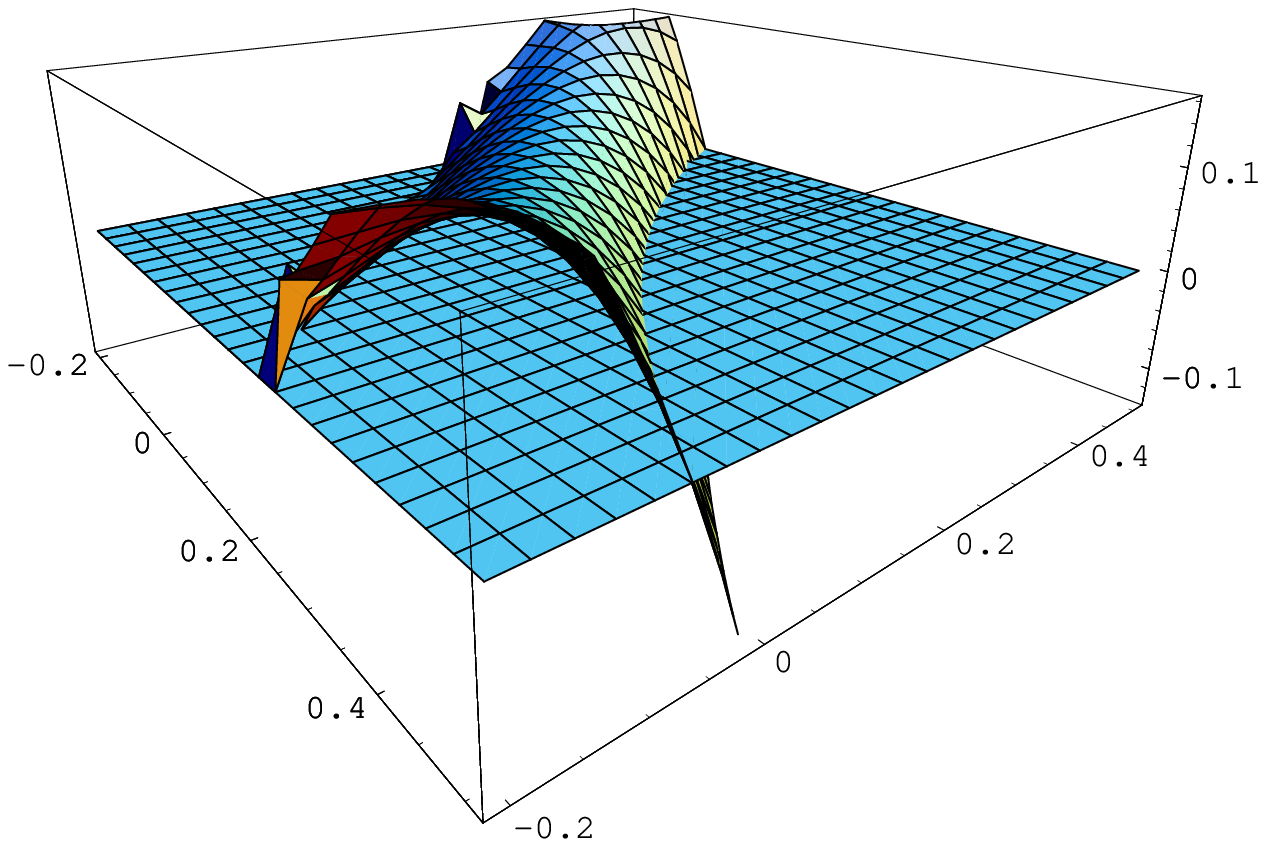}{Plot of the Euclidean action of the BPS BH as a
function of the parameters $(a,b)$. The flat plane corresponds
zero level surface.}
\noindent In the plot, it is easy to see that $I_{bps}$ is
positive for small $(a,b)$ and negative for larger values. From
which we deduce that there is a phase transition, where the BH
solution is not any more the preferred vacuum, but a meta-stable
vacuum. In (fig. 2) we show a two dimensional plot for $b=.1$
where the change of sign of $I_{bps}$ is more explicit. We are not
sure what is the stable vacuum, perhaps is one of the more general
solutions that still we do not know or perhaps is probably related
to a gas of superparticlas in AdS, studied in detail in
\cite{Kinney:2005ej}.
\EPSFIGURE[t]{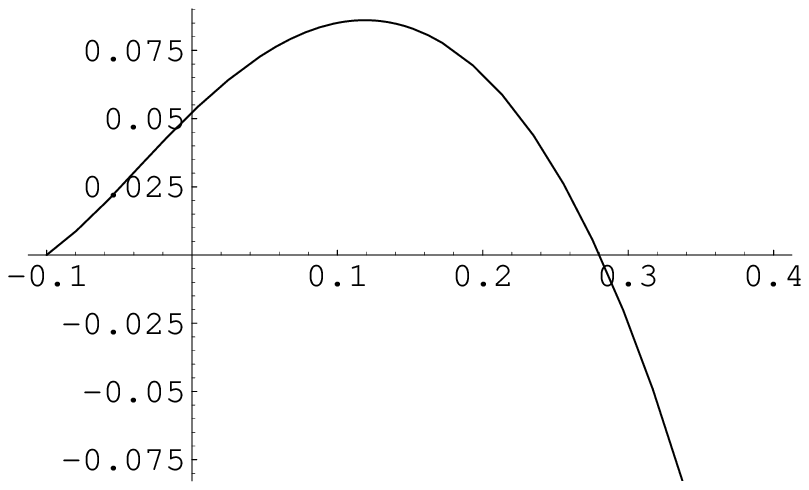,,scale=1.1}{Plot of the Euclidean action at
fixed $b=1/10$.}

After the above global analysis, we consider the local stability
criteria (see for example
\cite{Cai1,Cai2,Chamblin:1999tk,Chamblin:1999hg,Cvetic:1999rb,Cvetic:1999ne}),
base on the behavior of different susceptibilities that are
generalization alike the more traditional specific heat in the
canonical ensemble. There are many different ways introduce the
local stability analysis, but it can always be related to the
second law of thermodynamics where the entropy is a local maximum
of a stable equilibrium configuration. For example, we consider
the so-called "isothermal permittivity" $\ep$,
\bea \ep=-\left(\fft{\partial Q}{\partial
\phi}\right)_{T=0}\nonumber \eea
that relates the change of electric charge to the change of its
chemical potential $\phi$. In (fig. 3) we show a plot of $\ep$ as
a function of $(a,b)$, where it can be seen a first order
transition characterizing the phase transition. Also for more
clarity we show a two dimensional plot of $\ep$ at $b=.1$ as a
function of $a$, where the system is symmetric with respect to
$(a,b)$.

Other susceptibilities can be introduced, but we believe that the
above calculation illustrates well enough the thermodynamical
properties of the solution in the grand canonical ensemble.
\EPSFIGURE{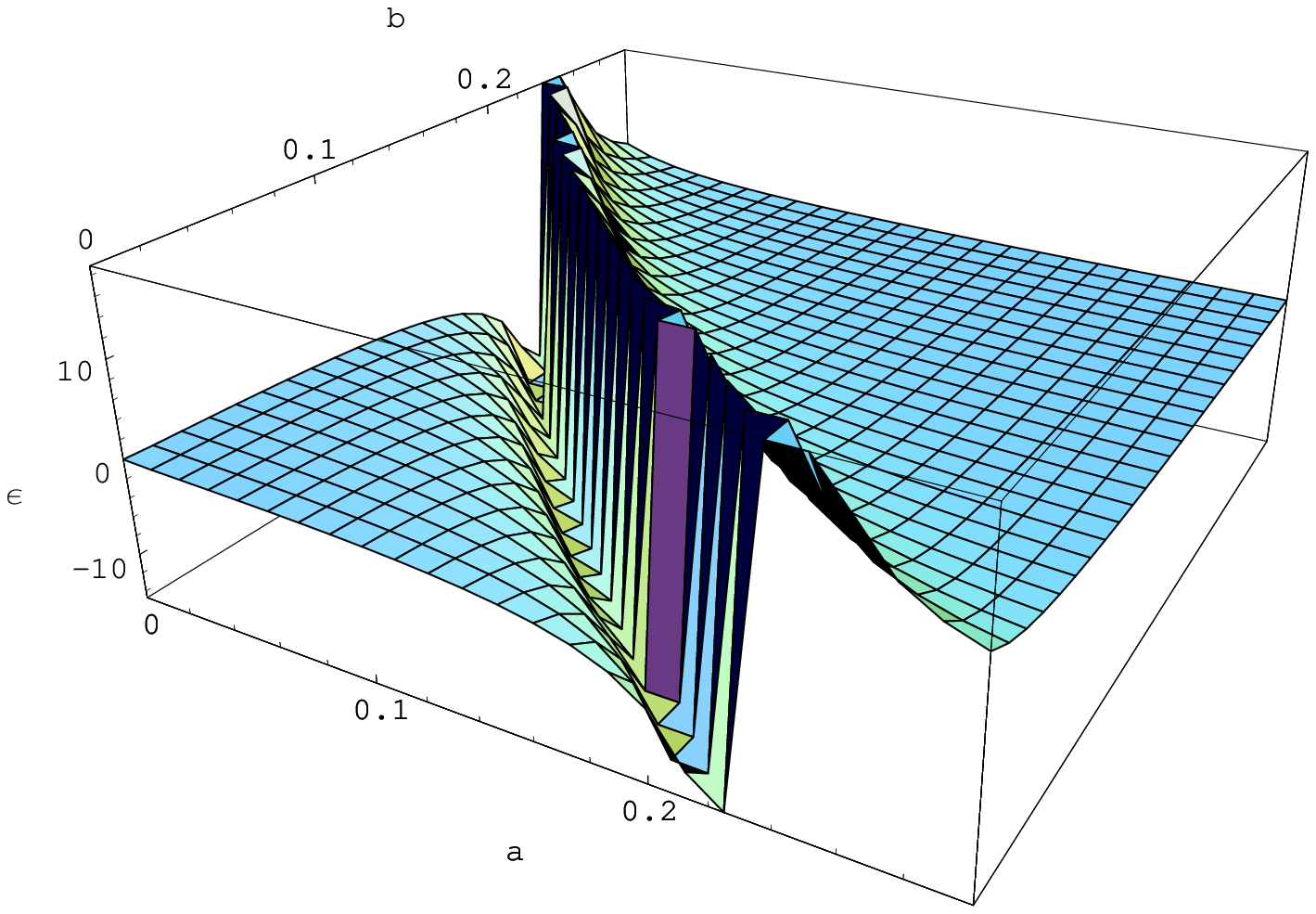,,scale=.75}{Plot of the susceptibility $\epsilon$
for $b=1/10$.}
\EPSFIGURE[h]{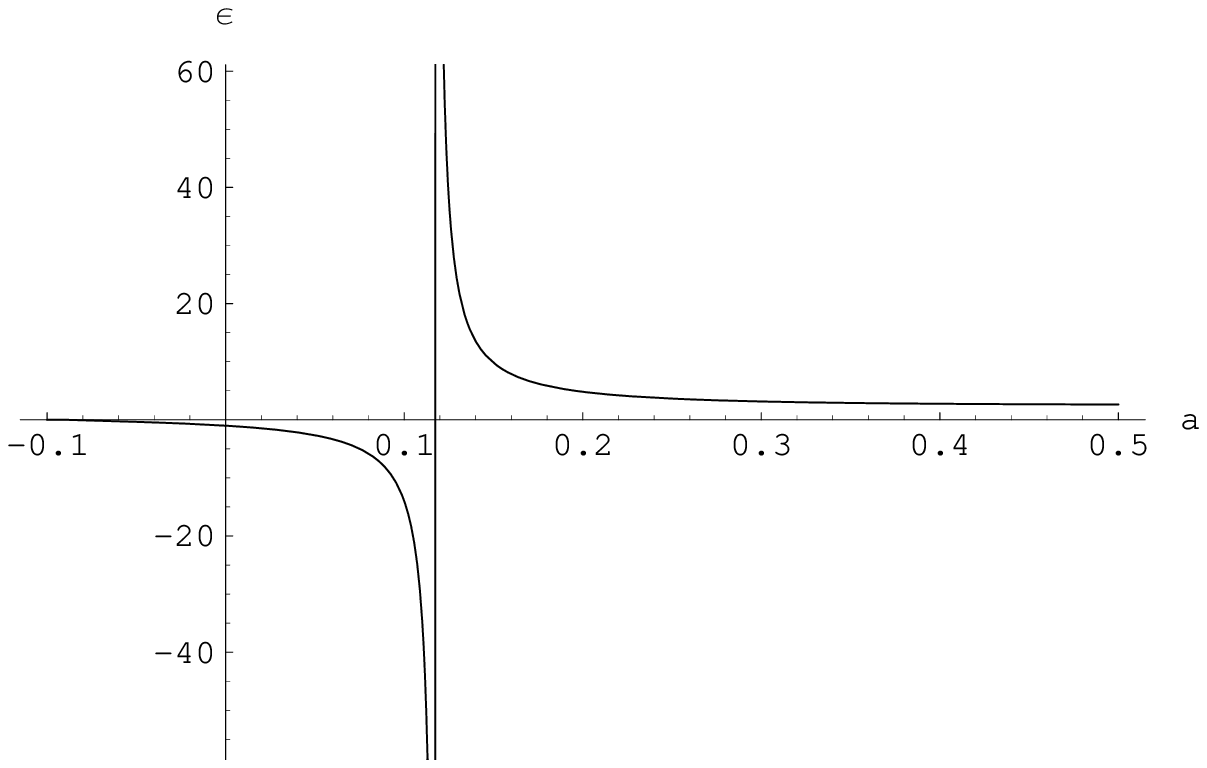,,scale=1}{Plot of the susceptibility $\epsilon$
for $b=1/10$.}

\subsection{\bf BPS solitonic solutions}\vspace{.3cm}

In this case, Evaluation of the $I_{bps}$ gives
\bea I_{bps}=
\fft{(b-1)^2(b^2+4b+1)^2(29b^5+232b^4+54b^3-174b^2-167b-46)}{6(b+1)
(2b+1)^5(b^4-6b^3+17b^2+16b+8)}\,,\nonumber \eea
where after some inspection, it is not difficult to see that among
all the possible ranges of $b$, the interval $\left((10-
\sqrt{139})/3,-1/2\right)$ covers all the physical possibilities.
 The other values of $b$, correspond either to repetitions of the
 relevant physical situation or produce solutions with
negative energy that we ruled out.

In figure (fig. 5) and (fig. 6), we show a plot of $I_{bps}$ as a
function of $(b)$ for this topological case. In the plots, it can
be seen that the $I_{bps}$ is positive in the first plot while
negative in the second. Therefore we again have a situation where
the solitonic solutions is unstable, and another where it is
stable.

We have found that at the point $b=-1/2$ the plot really shows a
break down of our expansion on the off-BPS parameter $\mu$.
Basically, at $b=-1/2$ we have done a "division by zero" and our
expression for $I_{bps}$ is not to be trust. Nevertheless, the
problem appears only at this very point, and we have checked that
it is possible to redefine a specific off-BPS expansion around
this particular point. Here we have not included such a detail
analysis since, in any case, the change of sign is warranty
because the expansion work fine away of $b=-1/2$.

The technical complications with the off-BPS expansion at the
point $b=-1/2$ prevent us to consider a local analysis of the
phase transition. This sort of studies will be covered in a
following work, where more extensive studies on the thermodynamic
properties of this and another family of solutions will be
reported \cite{futuro}.
\DOUBLEFIGURE{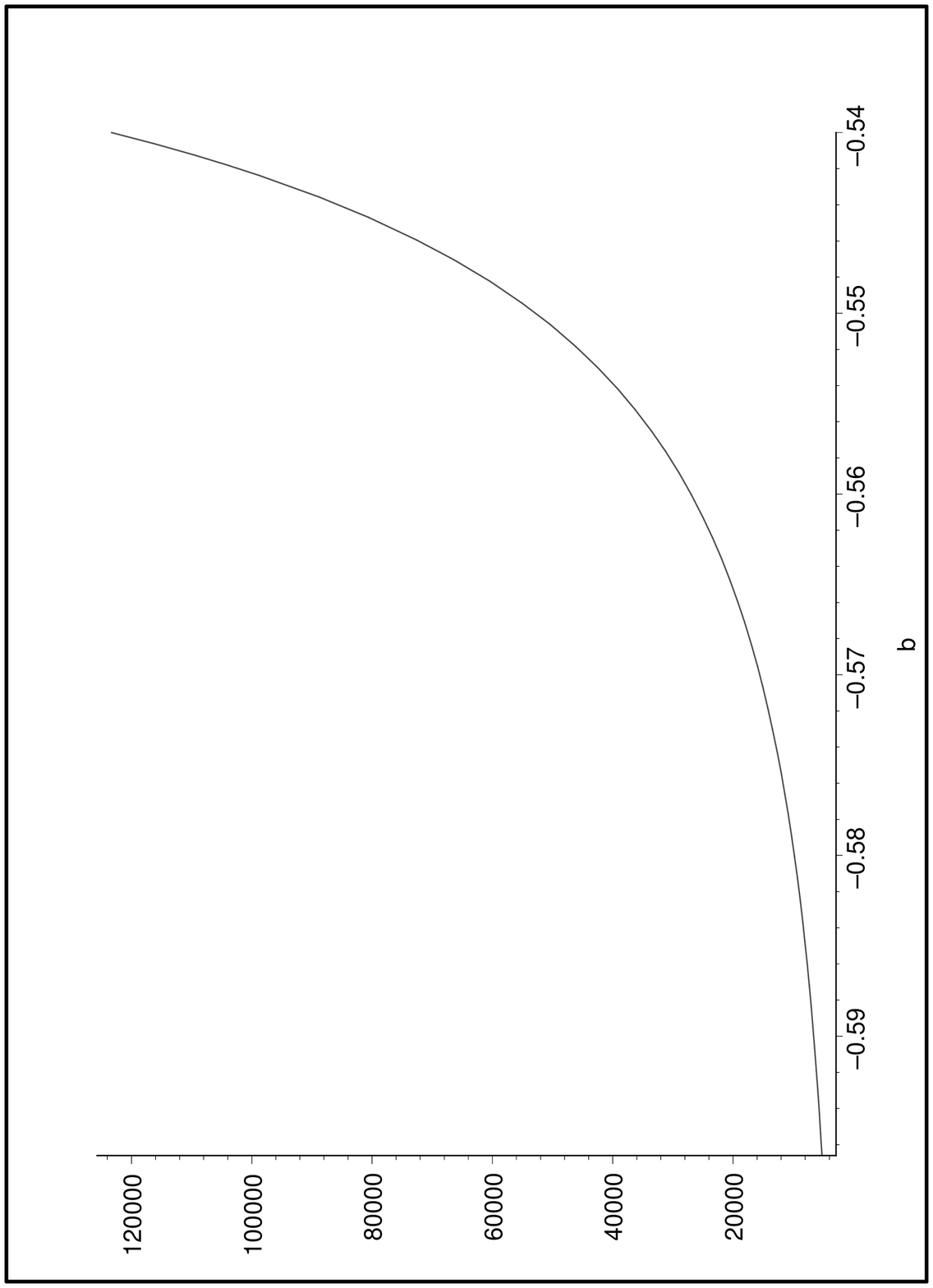,angle=-90,scale=.3}{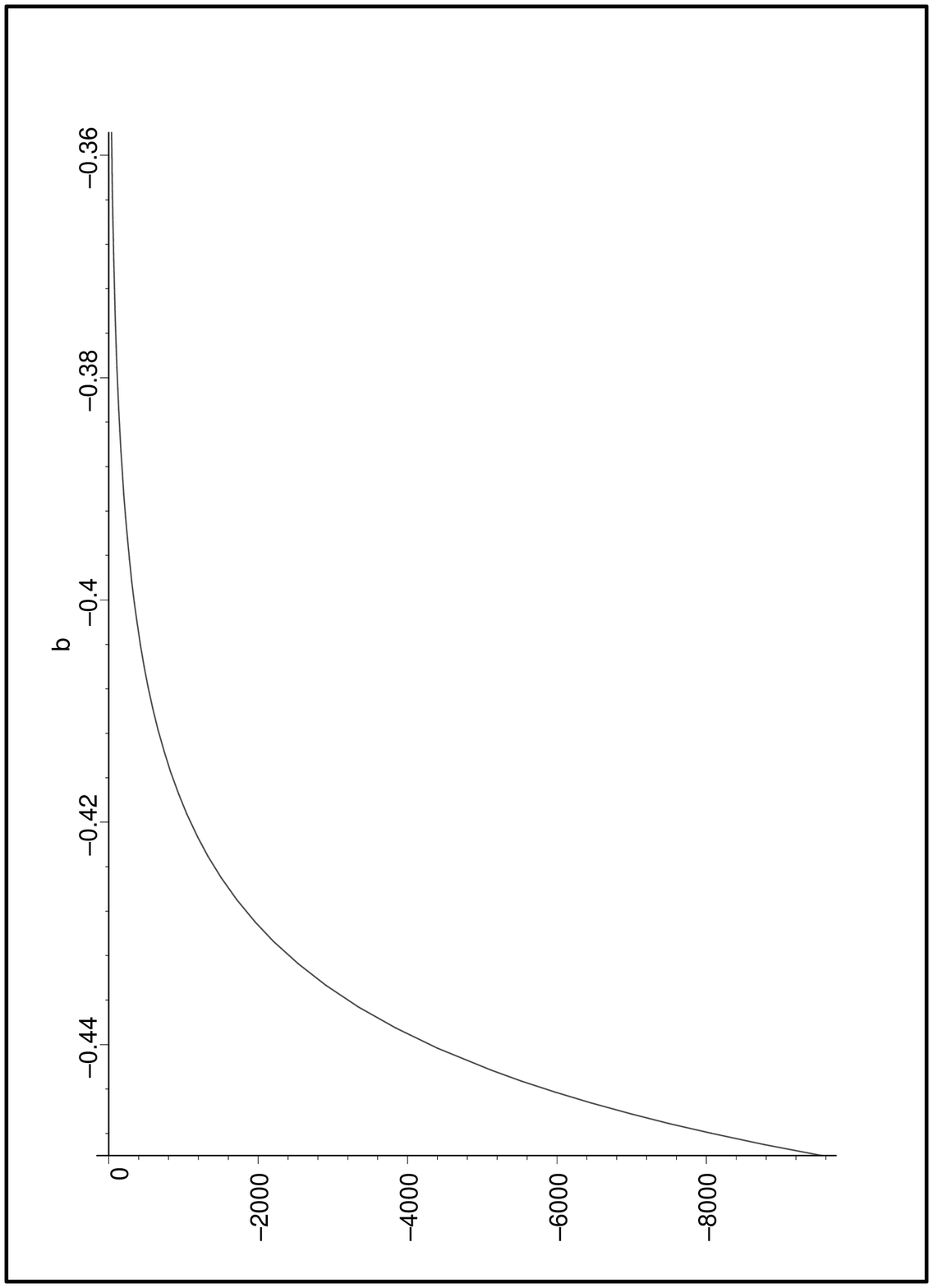,angle=-90,scale=.3}{This
plot cover the first half of the allowed range for $b$.}{This plot
cover the second half for the allowed range of $b$}

\section{Discussion}
\label{sec5}

In this work, we have defined a framework to study thermodynamics
properties in the BPS regime for general supersymmetric solitons
in gauge supergravity. The mechanism is suggested by the
equivalent more standard studies in supersymmetric field theories.
In particular due to the AdS/CFT duality, this kind of reasoning
acquires firmer grounds that support it.

To perform this analysis, we fund natural to assume that the
thermodynamic properties of the solitonic solutions in
supergravity are a fundamental characteristic that does not apply
only to solutions with non zero surface gravity and horizons like
the standard BH, but extends to supersymmetric solitons, with or
with out horizons at zero temperature. In other words,
\\

{\it the BH thermodynamics of space-time extends naturally to a
general principle of space-time physics}.
\\

In practice, how to extend the BH thermodynamics definitions is
not that clear. Here we found that things analogous to chemical
potentials to the different conserved charges do have an
extension. Again, the AdS/CFT duality also supports this
generalization since statistical mechanics is well defined in the
dual CFT theory and therefore should have a counterpart in the
supergravity side.

In fact, we have found explicit realizations of this mechanism and
framework in different types of supersymmetric solutions of
different supergravity theories, like the rotating BTZ BH in three
dimensions or the AdS BH in five dimensions. In all the examples
here studied, the limiting procedure gives finite quantities that
play the role of thermodynamical variables in the BPS regime.

In particular, we arrive to the definition of BPS chemical
potentials that where unknown up to now in the literature. For the
rotating BTZ BH, the BPS chemical potential corresponds exactly to
the inverse of the left temperature in the dual two dimensional
CFT, an encouraging signal that this BPS chemical potentials are
physical quantities that deserve attention. For the other studied
cases, we do not know their role in the CFT dual picture, so more
work is needed to address this important question.

We have also fund a rich structure of phase transitions among our
BPS examples. In particular the five dimensional soliton shows for
the BH case, a clear first order phase transition to another
soliton that is either a new solution that we do not know about or
simply a gas of superparticles in AdS. For the topological
solution, we found a instability, but where unable to determine
its order due to technical details.

The present work is just the initial step to study the BPS phase
space for BH in AdS. Here, we have defined the framework and
minimal machinery to obtain the phase diagrams. Then with a couple
of examples, we have checked that our initial assumptions hold,
and scan superficially the phase diagram of the solutions. We will
address in future works, a more detail and comprehensive study of
the phase diagram for this and other supersymmetric BH
\cite{futuro}.

We point out that in this work, we only worked examples that have
a well defined BPS limit, leaving untouched other interesting
solutions like the famous superstar \cite{Behrndt:1998ns}. In this
case, if you perform the off-BPS expansion it is easy to see that
the conserved charges and different potentials, do not obey the
scaling of eqn. \ref{limit1} and \ref{limit2}. Recalled that in
this solution the BPS regime is characterized by a singular
solution that is believed to receive string
corrections\cite{Suryanarayana:2004ig}. If this is true, and the
solution is corrected, the corresponding thermodynamical functions
will also receive modifications that in turns should produce the
correct limiting behaviour. We are wandering about the possibility
of reversing the above argument to find the form of the relevant
string corrections based on a well behaved thermodynamical limit.

There are of course, many interesting avenues that opens up at
this point, like the study of all BPS BH solutions, and not only
those with $1/16$ supersymmetry. In particular, since we do not
know how to define the chemical potentials in the BPS limit if we
do not know the off-BPS regime, it is very important to find that
corresponding off-BPS solutions to the strict solutions of
\cite{Kunduri:2006ek}. Also, in the CFT picture there are other
phase transitions, even in less supersymmetric sectors with $1/4$
or $1/8$ preserved supersymmetry, that would be very interesting
to address from the AdS point of view.


\section*{Acknowledgments}

\small The author would like to thanks the organizers of the
conference "Eurostrings 2006" where the beginning of this work
took place, due to stimulating conversations. In particular, we
thank J. Maldacena, for clarifications on the dual CFT statistical
mechanics approach. Also we thank D. klem, R. Emparan and J. Russo
for useful comments during this research.

This work was partially supported by INFN, MURST and by the
European Commission RTN program HPRN-CT-2000-00131, P.~J.~S. is
associated to the University of Milan and IFAE University of
Barcelona. \normalsize


\end{document}